\documentclass[mathleft
]{an}
\usepackage{graphicx}
\usepackage{times}
\usepackage{natbib}
\bibpunct{(}{)}{;}{a}{}{,}
\begin{document}

% The following seven commands are intended for editorial usage and should be ignored by
% the author(s).
\Pagespan{789}{}% Document's page range. 
% If second parameter is left empty, the last page is computed automatically.
\Yearpublication{2006}%
\Yearsubmission{2005}%
\Month{11}%   
\Volume{999}%  
\Issue{88}% 
% \DOI{This.is/not.aDOI}% 

\title{The extended minimum of solar cycle~23 as seen by radial velocity (GOLF, GONG) and intensity (VIRGO) helioseismic instruments}

\author{D. Salabert\inst{1,2}\fnmsep\thanks{Corresponding author:
  \email{salabert@iac.es}\newline}
\and  R.~A. Garc\'ia\inst{3}
\and  P.~L. Pall\'e\inst{1,2}
\and S.~J. Jim\'enez-Reyes\inst{1}
\and A. Jim\'enez\inst{1,2}
}
\titlerunning{The extended minimum of solar cycle~23 as seen by GOLF, GONG, and VIRGO}
\authorrunning{D. Salabert et al.}
\institute{Instituto de Astrof\'isica de Canarias,  E-38200 La Laguna, Tenerife, Spain
\and 
Departamento de Astrof\'isica, Universidad de La Laguna, E-38205 La Laguna, Tenerife, Spain
\and 
Laboratoire AIM, CEA/DSM-CNRS, Universit\'e Paris 7 Diderot, IRFU/SAp, Centre de Saclay, F-91191 Gif-sur-Yvette, France}

%\received{30 May 2005}
%\accepted{11 Nov 2005}
%\publonline{later}


\abstract{We present an analysis of the variability of the solar oscillation spectrum during solar cycle 23 and its extended minimum. We use simultaneous observations of the low-degree solar p modes collected by the space-based, Sun-as-a-star GOLF (radial velocity) and VIRGO (intensity) instruments, and by the ground-based, multi-site network GONG. We investigate in particular the response of the p-mode eigenfrequencies to the observed peculiar deep solar minimum of surface activity of 2007--2009 as compared with the previous solar cycle 23. We study the different temporal variations of the p-mode frequencies with individual angular degrees. }

\maketitle

\section{Introduction}
The {\it new millennium solar activity minimum} of 2007--2009 has shown the quietest Sun in almost a century with a delayed onset of solar cycle~24. Helioseismic observations ha\-ve been used to study the response of the solar oscillations to this unusually extended minimum of solar surface activity. While the temporal variations of the p-mode frequencies are closely correlated with solar surface activity proxies during the past cycles at low-  \citep[e.g.,][]{gelly02,salabert04} and higher-angular degrees \citep[e.g.,][]{chano01,salabert06a}, significant variations of the p-mode frequencies during the minimum of cycle~23 in contrast to the surface activity observations over the same period have been reported \citep{salabert09, broom09}. Furthermore, \citet{howe09} showed that the lack of sunspots and the low-activity levels during the cycle~23 minimum can be explained by a slower than usual jet stream associated to the production of the sunspots. These streams originating from the poles every 11 years migrate slowly below the surface towards the equator. 
Other p-mode parameters, such as the mode power and lifetime for instance, were also proven to be sensitive to the solar activity cycle in both Sun-as-a-star \citep[e.g.,][]{salabert03, chano04} and spatially-resolved \citep[e.g.,][]{komm00,salabert06b} observations. Thus, their response to the unusually long minimum of cycle~23 is also of great interest and is currently under investigation.

\begin{figure*}[t]
\centering
\includegraphics[scale=0.28]{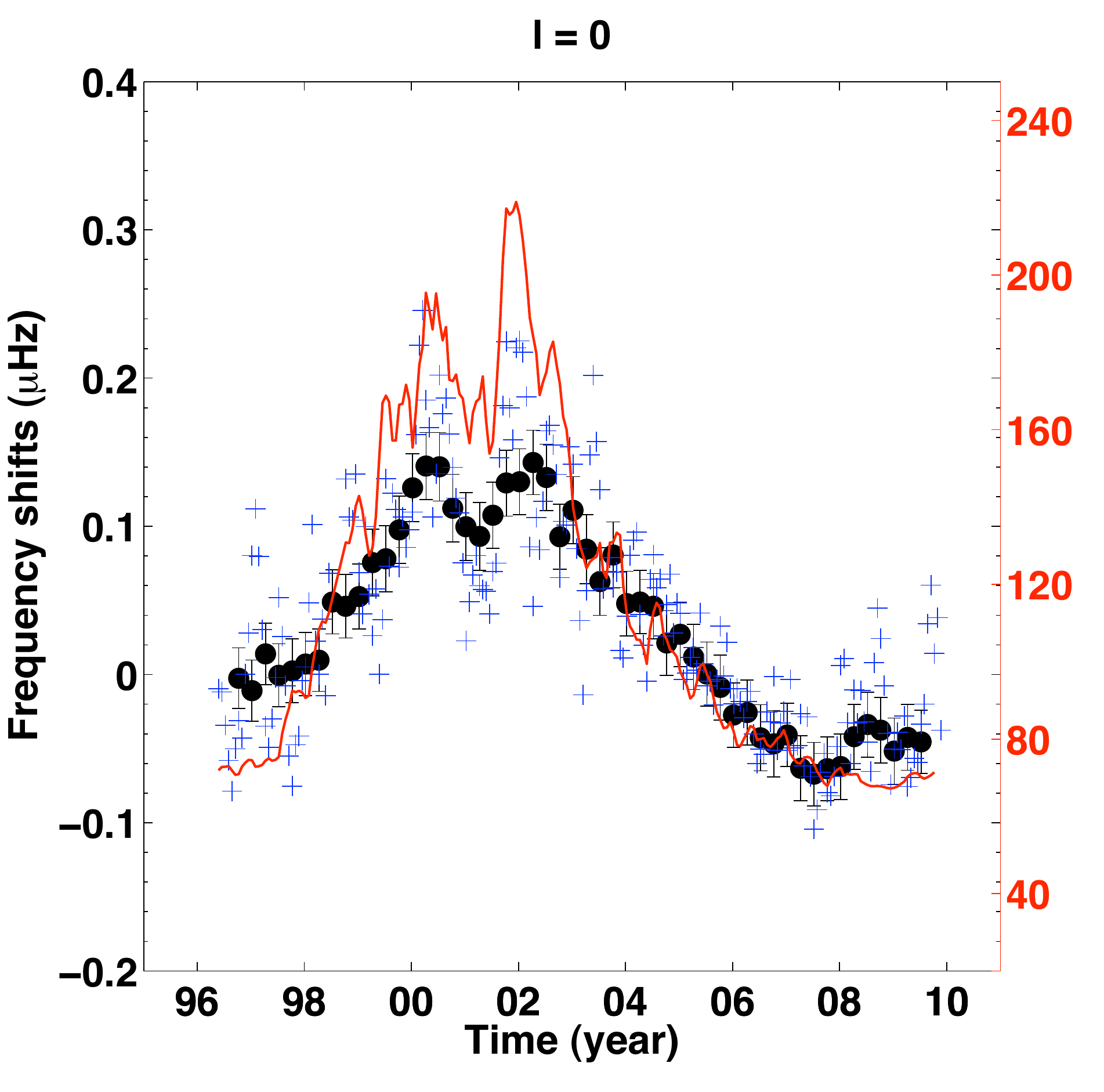}\includegraphics[scale=0.29]{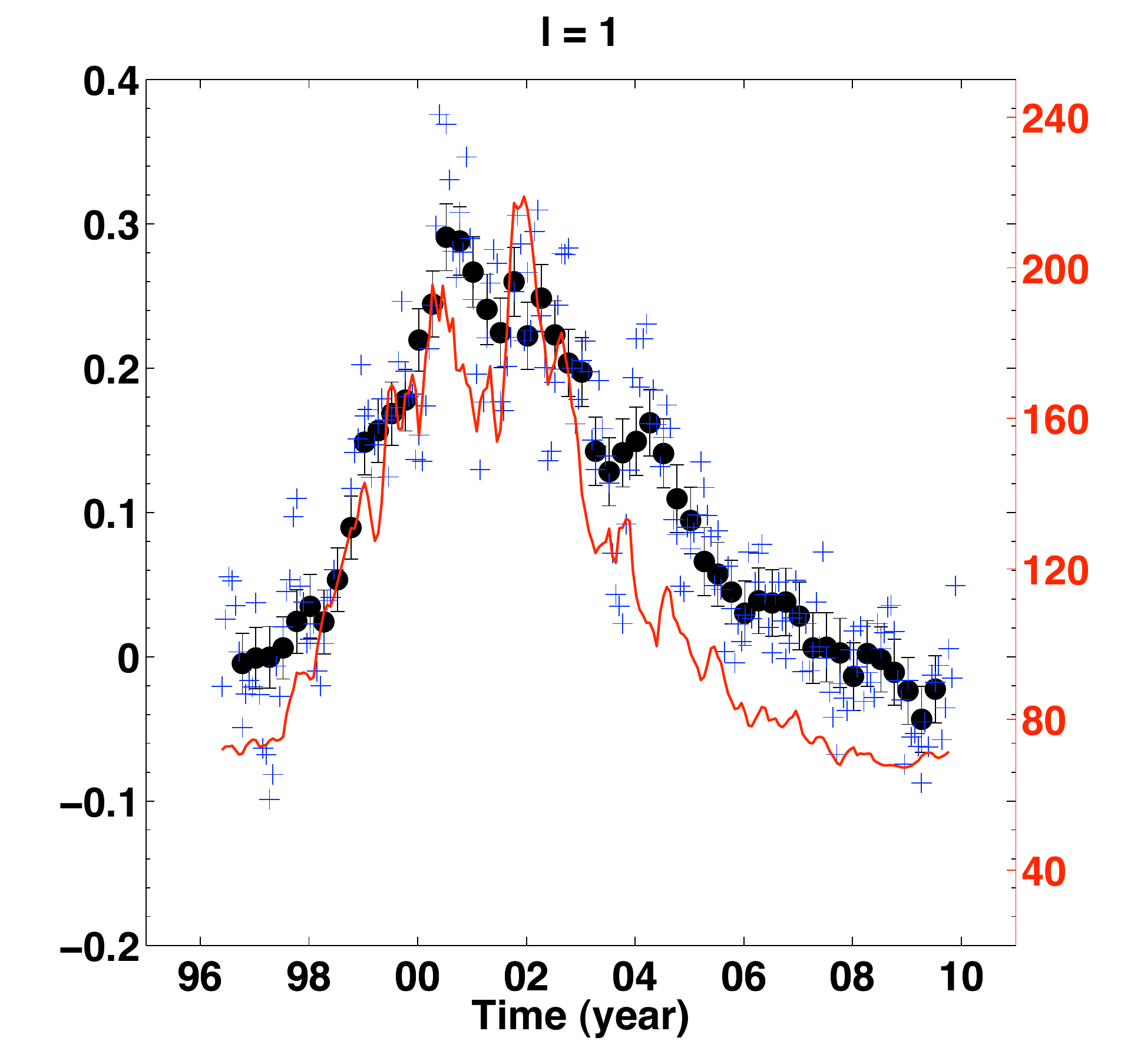}\includegraphics[scale=0.30]{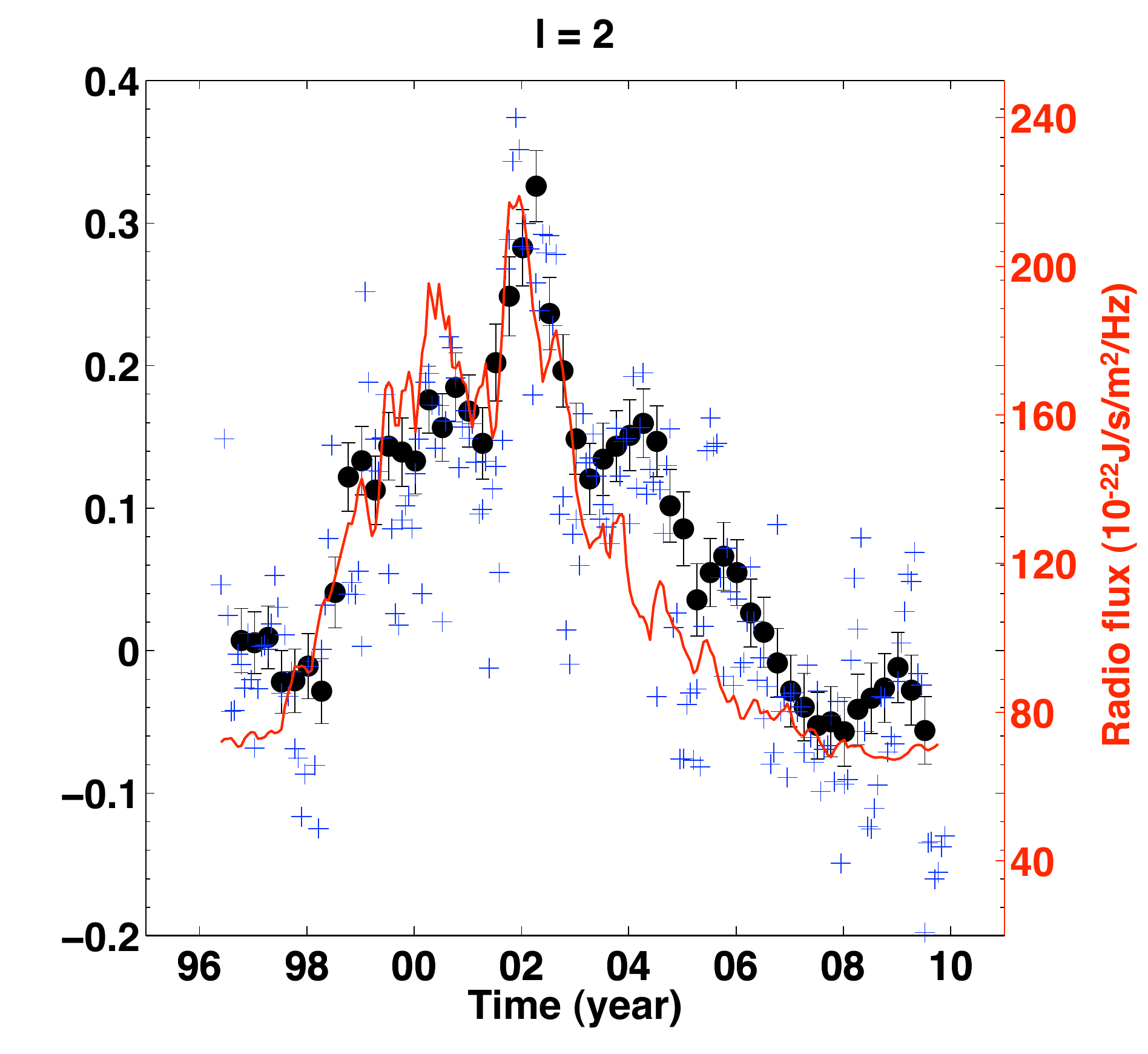}
\caption{Frequency shifts of the $l = 0, 1,$ and 2 solar p modes (left to right panels) extracted from the analysis of the 365-day (black dots) and the 91.25-day (blue plus signs) GOLF spectra. The associated error bars of the 365-day frequency shifts are represented. The corresponding 10.7-cm radio flux averaged over the same 91.25-day timespan is shown as a proxy of the solar surface activity (red solid line). }
\label{fig:fig1}
\end{figure*}

\begin{figure*}
\centering
\includegraphics[scale=0.75]{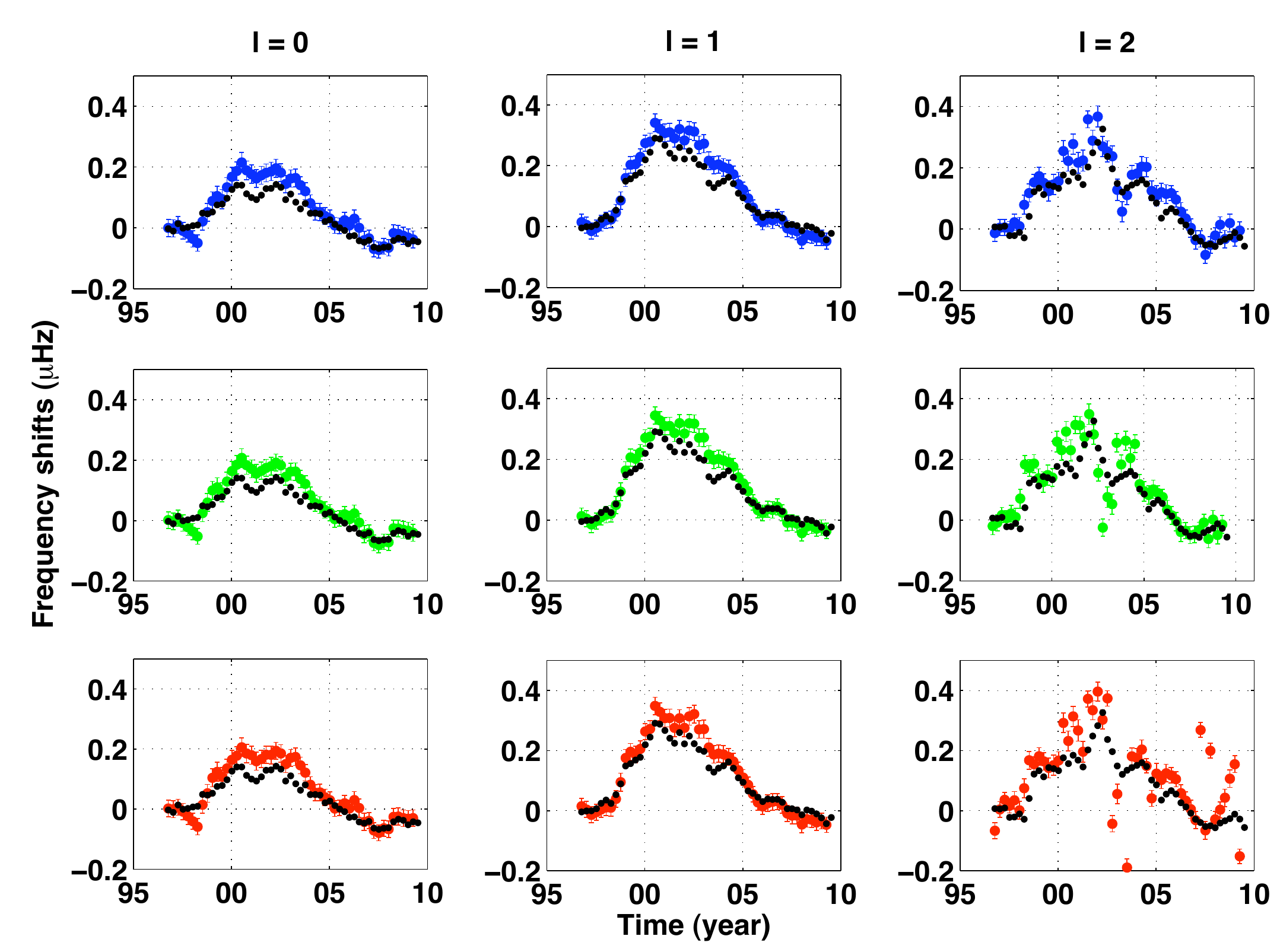}
\caption{Frequency shifts of the $l = 0, 1,$ and 2 solar p modes (left to right panels) extracted from the analysis of the 365-day VIRGO spectra. The frequency shifts measured from the blue, green, and red VIRGO 
channels are represented from top to bottom respectively. The corresponding GOLF frequency shifts are shown for comparison (black dots). The associated error bars are also represented.}
\label{fig:fig2}
\end{figure*}

\begin{figure*}
\centering
\includegraphics[scale=0.295]{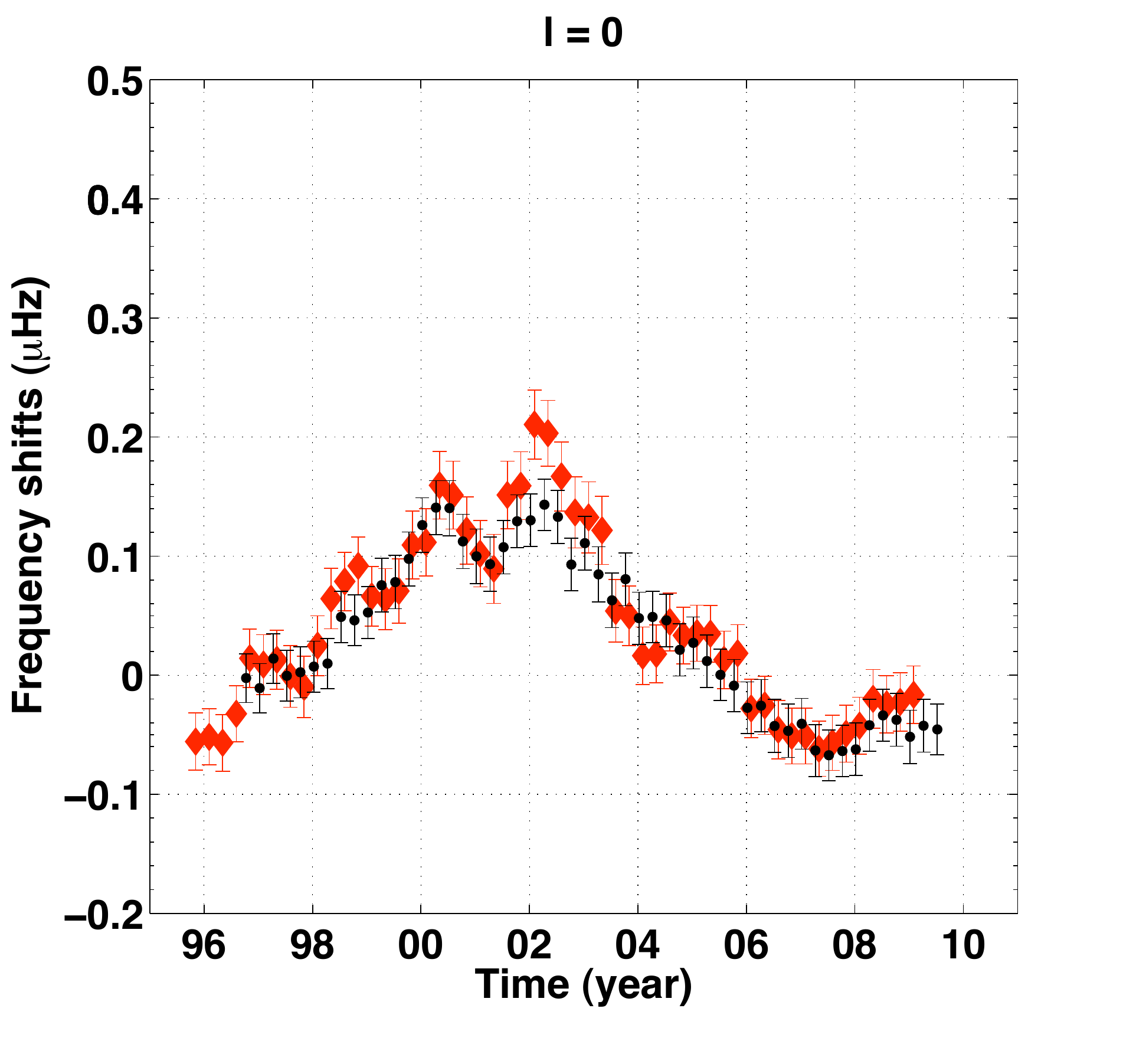}\includegraphics[scale=0.295]{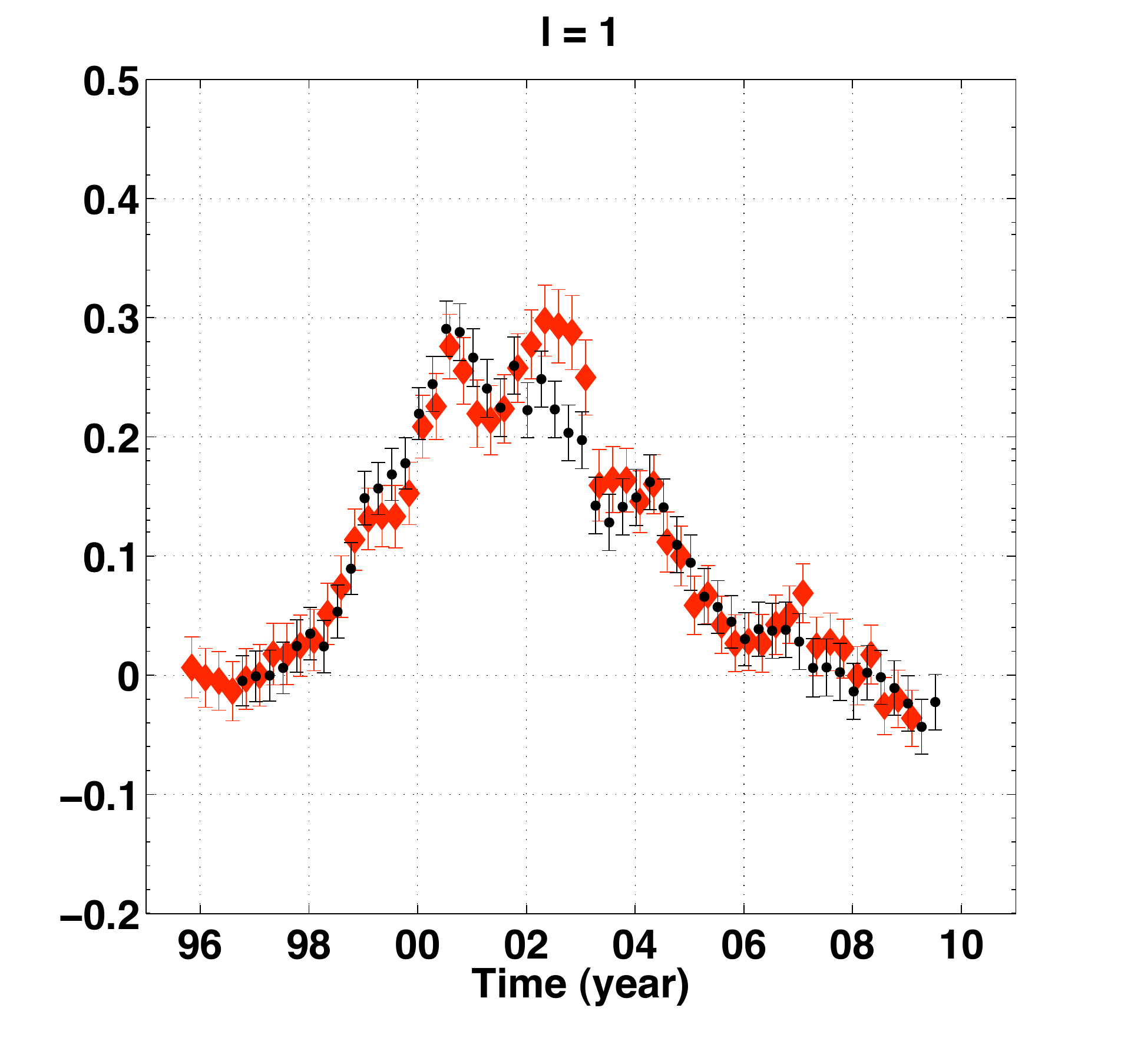}\includegraphics[scale=0.295]{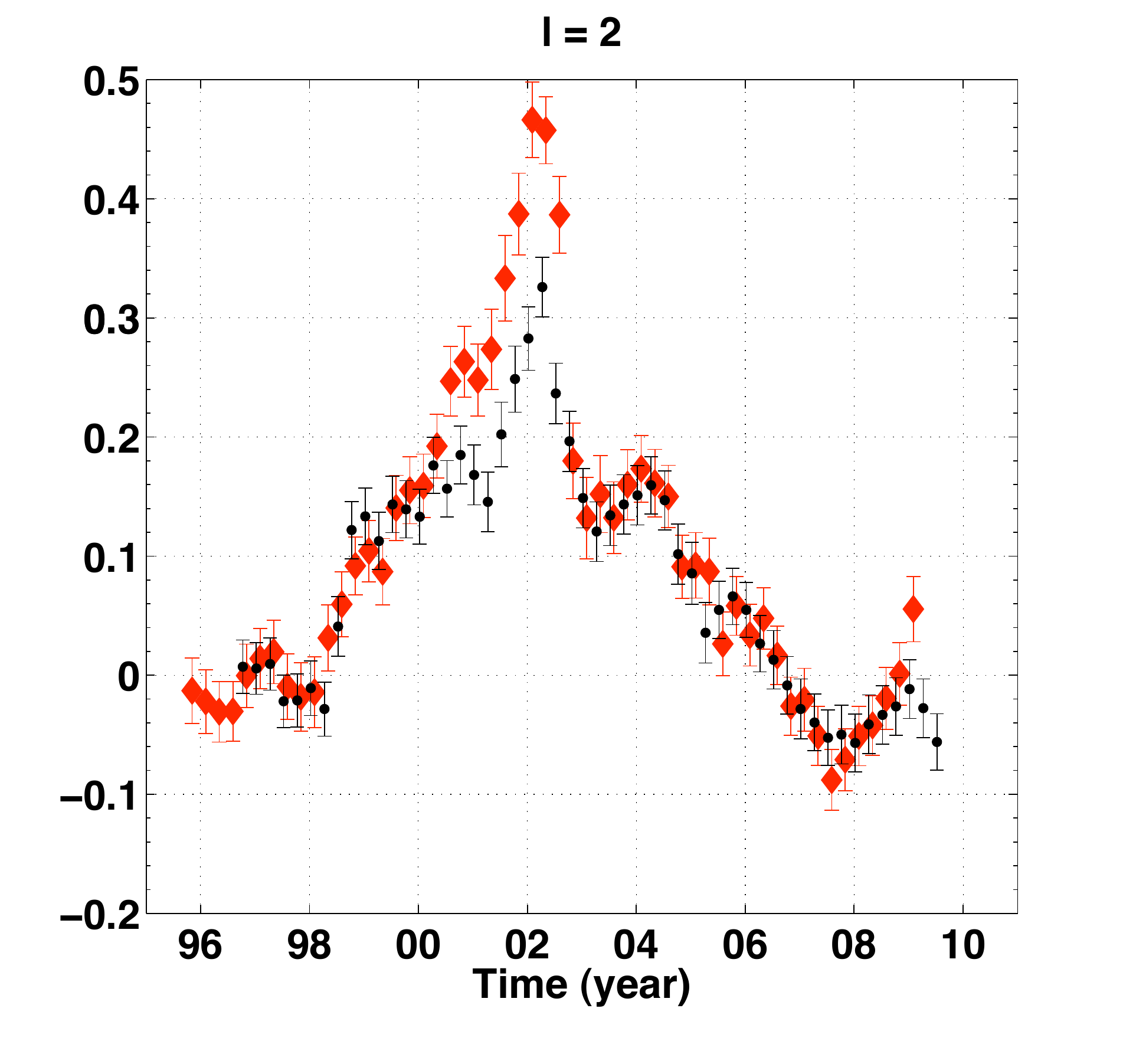}\caption{Frequency shifts of the $l = 0, 1,$ and 2 solar p modes (left to right panels) extracted from the analysis of the 365-day integrated GONG spectra (red diamonds). The corresponding GOLF frequency shifts are shown for comparison (black dots). The associated error bars are also represented. }
\label{fig:fig3}
\end{figure*}

\section{Observations and analysis}
Simultaneous helioseismic observations collected by three independent instruments are used in this work:
\begin{itemize}
\item the space-based Global Oscillations at Low Frequency \citep[GOLF;][]{gabriel95} on board the {\it Solar and Heliospheric Observatory (SoHO)} spacecraft. GOLF is a resonant scattering spectrophotometer. It measures the Doppler wavelength shift -- integrated over the solar surface -- in the D1 and D2 Fraunhofer sodium lines at 589.6 and 589.0~nm respectively;
\item the space-based instrument Variability of Solar IRrandiance and Gravity Oscillation \citep[VIRGO;][]{froh95} instrument on board {\it SoHO}. VIRGO is composed of three Sun photometers (SPM) at 402~nm (blue channel), 500~nm (green channel), and 862~nm (red channel);
\item the ground-based, multi-site instruments Global Oscillation Net\-work Group \citep[GONG;][]{harvey96} is composed of 6 identical stations located at selected longitudes around the world. The GONG instruments are Michelson Doppler Interferometers measuring in the absorption line Ni I (676.8~nm).
\end{itemize}

These datasets were split into contiguous 365-day and 91.25-day sub-series, each series being allowed to overlap by 91.25 days and 22.8125 days respectively. The power spectrum of each sub series was fitted to estimate the mode parameters \citep{salabert07} using a standard likelihood maximization function (i.e. power spectrum with a $\chi^2$ with 2 d.o.f. statistics). Each mode component is parameterized using a asymmetric Lorentzian profile. The temporal variations of the p-mode frequencies were defined as the difference between reference values (taken as the average over the years 1996--1997) and the frequencies of the corresponding modes observed at different dates. The weighted averages of these frequency shifts were then calculated between 2200 and 3300~$\mu$Hz. Mean values of daily measurements of the 10.7-cm radio flux were obtained and used as a proxy of the solar surface activity.

\subsection{Space-based radial velocity GOLF/SoHO}
A total of 5021 days of radial velocity GOLF time series \citep{Ulrich00,Garcia05} were analyzed. This dataset spans the period from 1996 April 11 to 2010 January 8, with a overall duty cycle of 95.3\% (see \citet{chano03} for the calibration method). The frequency shifts measured at each angular degree ($l = 0, 1,$ and 2) in the 91.25-day and 365-day sub series are shown on Fig.~\ref{fig:fig1}. The corresponding 10.7-cm radio flux averaged over the same 91.25-day timespan is also represented as a proxy of the solar surface activity.

\subsection{Space-based photometer VIRGO/SoHO}
We also analyzed 4890 days of the intensity VIRGO time series. This dataset starts on 1996 April 11 and ends on 2009 August 30, with a  duty cycle of 94.6\%. The 365-day frequency shifts at $l = 0, 1,$ and 2 measured in the blue, green, and red VIRGO channels are represented on Fig.~\ref{fig:fig2}. The 365-day GOLF frequency shifts are also shown in black for comparison.

\subsection{Ground-based, multi-site GONG}
A total of 5112 days of the integrated time series of the ground-based, multi-site GONG network spanning the period from 1995 May 5 to 2009 May 4 were analyzed, with an overall duty cycle of 85.4\%. Due to the spatial resolution of the original GONG data, part of the power from the higher angular degrees ($l \geq 4$) are present in the integrated GONG signal. These leaks were taken into account during the peak-fitting by including information coming from the GONG leakage matrix.
The frequency shifts at $l = 0, 1,$ and 2 measured in the 365-day integrated GONG sub series are represented on Fig.~\ref{fig:fig3}. The 365-day GOLF frequency shifts are also shown.

\section{Results}
 The solar p-mode frequencies measured simultaneously  by independent space-based and ground-based helioseismic instruments -- GOLF and GONG/integrated signal in radial velocity, and VIRGO in intensity -- show similar temporal variations during solar cycle 23 and its extended minimum of 2007--2009. Moreover, and as shown on Figs.~\ref{fig:fig1},~\ref{fig:fig2}, and \ref{fig:fig3}, different behaviors are observed amongst modes of different angular degrees ($l=0,1$, and 2):
 \begin{itemize}
 \item the $l = 0$ and $l = 2$ frequency shifts show an upturn from the end of 2007 while no significant surface activity was visible on the Sun. This upturn might be followed by a downturn after 2009;
 \item the $l = 1$ frequency shifts follow the general decreasing trend of the solar surface activity as clearly illustrated on Fig.~\ref{fig:fig1}.
 \end{itemize}
 
It is worth noticing the particular behavior of the $l=2$ frequencies in the VIRGO red channel data. While, the $l=2$ mode is overall noisier in the VIRGO observations than in the GOLF and GONG data, the sharp increase of the $l=2$ frequency shifts after 2007 in the VIRGO  red channel observations is striking and does not seem to be only due to a higher noise level. However, more work is needed to conclude on the significance of the different sharpness of these upturns, for example to determine if there is a depth dependence of the observed upturn since 2007.\\

Nevertheless, the differences between individual angular degrees might be interpreted as different geometrical responses to the spatial distribution of the solar magnetic field beneath the surface of the Sun. That could indicate variations in the magnetic flux at high latitudes related to the onset of solar cycle 24.

\section{Conclusions}
We analyzed simultaneous observations of the low-degree solar p modes collected by the space-based GOLF (radial velocity) and VIRGO (intensity) instruments on board the {\it SoHO} spacecraft, and by the ground-based, multi-site network GONG in order to investigate the response of the p-mode frequencies to the unusual deep and long minimum of solar surface activity of cycle 23 during 2007--2009. We observed different temporal variations of the p-mode frequencies between individual angular degrees. These variations are identical in the GOLF, VIRGO, and GONG instruments. The differences between individual angular degrees might be interpreted as different geometrical responses to the spatial distribution of the solar magnetic field beneath the surface of the Sun. A more detailed intercomparison between several helioseismic instruments is underway.

After CoRoT revealed a first stellar activity cycle in a Sun-like star using asteroseismology \citep{garcia10}, the NASA {\it Kepler} mission \citep{koch10} will reinforce the study of stellar magnetic cycles \citep{karoff09} thanks to long-duration observations (over 3 years) with a very high signal-to-noise ratio \citep[e.g.,][]{chaplin10,bedding10}. Thus, seismology will provide very detailed inferences of the stellar structures, in particular the depth of the convective zones. Such information will be of key importance to properly understand the physical mechanisms driving the magnetic cycles and thus, will help to better understand and predict the dynamo processes taking place in the Sun.

\acknowledgements   
The GOLF and VIRGO instruments on board SoHO are a cooperative effort of many individuals, to whom we are indebted. SoHO is a project of international collaboration between ESA and NASA. This work utilizes data obtained by the Global Oscillation Network Group (GONG) program, managed by the National Solar Observatory, which is operated by AURA, Inc. under a cooperative agreement with the National Science Foundation. The data were acquired by instruments operated by the Big Bear Solar Observatory, High Altitude Observatory, Learmonth Solar Observatory, Udaipur Solar Observatory, Instituto de Astrof\'isica de Canarias, and Cerro Tololo Interamerican Observatory. The 10.7-cm radio flux data were obtained from the National Geophysical Data Center. D.S. acknowledges the support from the Spanish National Research Plan (grant PNAyA2007-62650). This work has been partially supported by the European Helio- and Asteroseismology Network (HELAS) and by the CNES/GOLF grant at SAp CEA-Saclay.

%%%%%%%%%%%%%%%%%%%%%%%%%%%%%%%%%%%%%%%%%%%%%%%%%%%%%%

\end{document}